\documentclass[sigconf,natbib=true,anonymous=false]{acmart}
\usepackage{booktabs}
\usepackage{multirow}
\usepackage{graphicx} 
\usepackage{enumitem}
\usepackage{subcaption}
\usepackage{array}
\usepackage{booktabs}
\AtBeginDocument{%
  }

\setcopyright{acmlicensed}
\copyrightyear{2018}
\acmYear{2018}
\acmDOI{XXXXXXX.XXXXXXX}
\acmConference[Conference acronym 'XX]{Make sure to enter the correct
  conference title from your rights confirmation email}{June 03--05,
  2018}{Woodstock, NY}
\acmISBN{978-1-4503-XXXX-X/2018/06}

\begin{document}

\title{Drift-Aware Continual Tokenization for Generative Recommendation}

\author{Yuebo Feng}
\orcid{0009-0006-1595-9881}
\affiliation{
  \institution{Fudan University}
  \city{Shanghai}
  \country{China}}
\email{ybfeng25@m.fudan.edu.cn}

\author{Jiahao Liu}
\orcid{0000-0002-5654-5902}
\affiliation{
  \institution{Fudan University}
  \city{Shanghai}
  \country{China}}
\email{jiahaoliu21@m.fudan.edu.cn}

\author{Mingzhe Han}
\orcid{0000-0002-4911-6093}
\affiliation{
  \institution{Fudan University}
  \city{Shanghai}
  \country{China}}
\email{mzhan22@m.fudan.edu.cn}

\author{Dongsheng Li}
\orcid{0000-0003-3103-8442}
\affiliation{
  \institution{Microsoft Research Asia}
  \city{Shanghai}
  \country{China}}
\email{dongshengli@fudan.edu.cn}

\author{Hansu Gu}
\orcid{0000-0002-1426-3210}
\affiliation{
  \institution{Independent}
  \city{Seattle}
  \country{United States}}
\email{hansug@acm.org}

\author{Peng Zhang}
\orcid{0000-0002-9109-4625}
\affiliation{
  \institution{Fudan University}
  \city{Shanghai}
  \country{China}}
\email{zhangpeng\_@fudan.edu.cn}

\author{Tun Lu}
\orcid{0000-0002-6633-4826}
\affiliation{
  \institution{Fudan University}
  \city{Shanghai}
  \country{China}}
\email{lutun@fudan.edu.cn}

\author{Ning Gu}
\orcid{0000-0002-2915-974X}
\affiliation{
  \institution{Fudan University}
  \city{Shanghai}
  \country{China}}
\email{ninggu@fudan.edu.cn}

\renewcommand{\shortauthors}{Trovato et al.}

\begin{abstract}
Generative recommendation commonly adopts a two-stage pipeline in which a learnable tokenizer maps items to discrete token sequences (i.e. identifiers) and an autoregressive generative recommender model (GRM) performs prediction based on these identifiers. Recent tokenizers further incorporate collaborative signals so that items with similar user-behavior patterns receive similar codes, substantially improving recommendation quality. However, real-world environments evolve continuously: new items cause identifier collision and shifts, while new interactions induce collaborative drift in existing items (e.g., changing co-occurrence patterns and popularity). Fully retraining both tokenizer and GRM is often prohibitively expensive, yet naively fine-tuning the tokenizer can alter token sequences for the majority of existing items, undermining the GRM’s learned token–embedding alignment. To balance plasticity and stability for collaborative tokenizers, we propose \textbf{DACT}, a \underline{D}rift-\underline{A}ware \underline{C}ontinual \underline{T}okenization framework with two stages: (i) tokenizer fine-tuning, augmented with a jointly trained Collaborative Drift Identification Module (CDIM) that outputs item-level drift confidence and enables differentiated optimization for drifting and stationary items; and (ii) hierarchical code reassignment using a relaxed-to-strict strategy to update token sequences while limiting unnecessary changes. Experiments on three real-world datasets with two representative GRMs show that DACT consistently achieves better performance than baselines, demonstrating effective adaptation to collaborative evolution with reduced disruption to prior knowledge. Our implementation is publicly available at \url{https://github.com/HomesAmaranta/DACT} for reproducibility.

\end{abstract}



\received{20 February 2007}
\received[revised]{12 March 2009}
\received[accepted]{5 June 2009}

\maketitle
\section{Introduction}
In recent years, generative recommendation~\cite{deng2025onerec,wang2023generative} has emerged as a new paradigm that reformulates recommendation as a sequence generation task, different from traditional discriminative methods~\cite{xia2022fire,liu2022parameter,liu2023personalized,liu2023triple}. Most approaches follow a two-stage design: a learnable tokenizer (e.g., RQ-VAE) maps items into discrete hierarchical token sequences as identifiers~\cite{rajput2023recommender,tong2026rq,han2026feature}, and then Generative Recommender Model (GRM) performs autoregressive generation and prediction based on these tokens~\cite{liu2025improving}. In contrast to constructing tokens solely from static content (e.g., titles, descriptions, photos), recent studies incorporate collaborative signals from collaborative filtering (CF) models into the tokenizer~\cite{wang2024content,wang2024learnable,wang2024eager,zhou2025onerec,Wang2025}. This ensures that items sharing similar collaborative signals in user behaviors are assigned similar token sequences~\cite{wang2024learnable}. This enables the tokenizer to capture both semantic information and collaborative relationship, leading to substantial performance gains in generative recommendation.

\begin{figure}[t]
    \centering
    \includegraphics[width=\columnwidth]{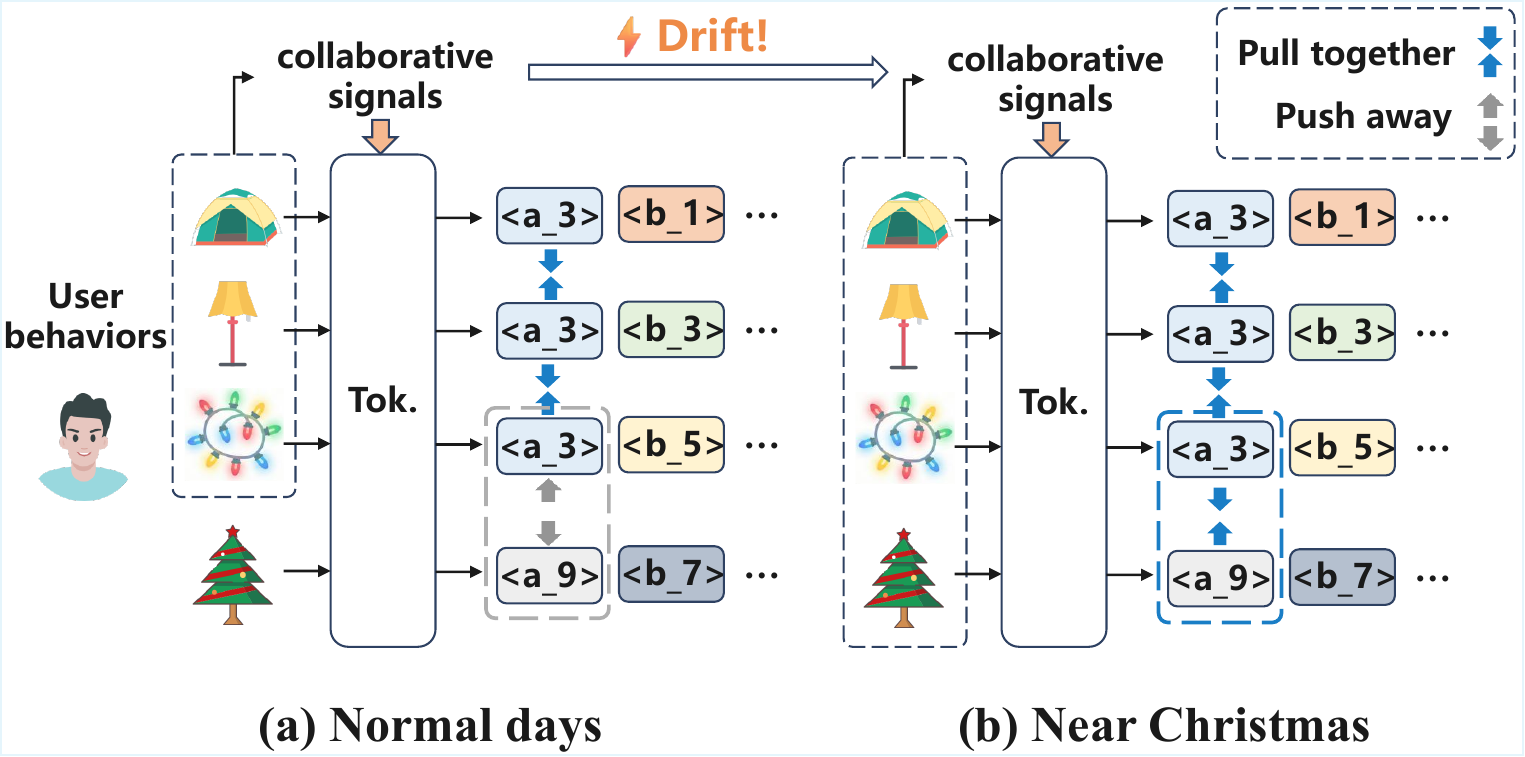}
    \caption{An example of item popularity and co-occurrence drift over time.}
    \label{fig:intro2}
\end{figure}

However, in real-world systems, with numerous new items and interactions arriving continuously, two challenges naturally arise: (1) During training, the tokenizer has not observed new items, which leads to identifier collision and shifts~\cite{shi2025incremental,wu2025bidirectional,liu2026distribution}, consequently leading to a decline in the performance of generative recommend model. (2) New interactions reflect underlying changes, including shifts in item popularity and evolving item co-occurrence patterns~\cite{koren2009collaborative,hidasi2015session,han2025fedcia}. If tokens are not adaptively updated to reflect such collaborative drift, outdated identifier will fail to represent the latest collaborative characteristics of items, becoming a key bottleneck for GRM performance~\cite{liu2025generative,qu2025tokenrec,gu2026llm}. As illustrated in Figure~\ref{fig:intro2}, in normal days, LED light strings co-occur more frequently with items such as floor lamps and tents, and thus share some codes. When Christmas approaches, LED light strings are widely used to decorate Christmas trees, and Christmas trees shift from long-tail items to popular ones, indicating collaborative signal drift. In this case, if the identifiers stay unchanged, it will be hard for GRM to capture the emerging co-occurrence pattern between LED light strings and Christmas trees, causing the Christmas trees to be under-ranked.

Although jointly retraining the tokenizer and the generative recommendation model (GRM) can resolve the mismatch caused by collaborative evolution, such a solution is often impractical due to its prohibitive computational cost and training latency~\cite{yoo2025embracing}. A more natural alternative is to incrementally fine-tune the tokenizer to absorb new collaborative signals. However, finetuning the tokenizer actually leads to more than 90\% of existing items experience changes in their token sequences~\cite{shi2025incremental}. Such drastic alteration invalidates the learned semantic representations from the previous stage, as token embeddings constitute the vast majority of parameters in industrial recommender systems~\cite{liu2021learnable}. 

From the perspective of continual learning, we are facing the well-known dilemma of how to balance learning plasticity (adapting to new knowledge) and memory stability (retaining past knowledge)~\cite{wang2024comprehensive,liu2025unbiased}. Recent works have begun to investigate this problem within the recommendation domain\cite{yoo2025continual,yoo2025embracing}. For GRM, introducing dedicated modules enables the model to capture both long- and short-term user preferences~\cite{yoo2025continual,shi2024preliminary}; for the learnable tokenizer, incremental updates typically assign tokens to new items while keeping old-item codes fixed for stability based on their static semantic information~\cite{shi2025incremental,zhang2025evalagent}. However, increasing works has introduced collaborative signals into the tokenizer~\cite{wang2024learnable, zhou2025onerec, wang2024eager}, which makes previous incremental methods inadequate, as they typically operate under the assumption of static item semantics, failing to capture the collaborative drift of existing items induced by evolving user behaviors. Meanwhile, the quality and alignment of item token sequences serve as the fundamental guarantee for GRM's recommendation effectiveness, acting as a prerequisite for performance improvement~\cite{liu2025generative,liu2025discrec,liu2025filtering}.

Consequently, the core research problem addressed in this paper is: in the incremental learning of collaborative-aware tokenizers, how can we accurately identify the subset of items undergoing significant collaborative semantic drift, and design a mechanism to selectively adapt their codes while strictly constraining stationary items? We term this a drift-aware selective adaptation strategy, which aims to achieve plasticity by adapting to evolving collaborative signals, while simultaneously ensuring the stability of the GRM's token embedding knowledge.

To this end, we propose \textbf{DACT}, a {D}rift-{A}ware {C}ontinual {T}okenization framework for generative recommendation incorporating collaborative signal. DACT captures and adapts to collaborative drift through two stages. In the first stage, we fine-tune the tokenizer to update the latent representations of each item. To accurately identify drifting items, we introduce a \textbf{C}ollaborative \textbf{D}rift \textbf{I}dentification \textbf{M}odule (CDIM) which is jointly trained with the tokenizer in an end-to-end manner. By learning drift patterns, this module outputs a drift confidence for each item. Based on this confidence, we adopt a differentiated training strategy: for drifting items, we encourage them to align with the latest collaborative signals; for stationary items, we apply additional constraints that anchor the current representations to those learned in the previous stage. In the second stage, we perform hierarchical code reassignment. Specifically, based on the updated latent representations, we employ a relaxed-to-strict hierarchical reassignment strategy to update the token sequence for each item, effectively balancing stability and plasticity.

Overall, the key advantage of our framework is its drift-aware selective strategy. By dynamically balancing plasticity and stability, it adapts to collaborative evolution efficiently, while preserving the alignment between item tokens and the GRM’s learned embedding space.

Our main contributions are summarized as follows:
\begin{itemize}
    \item We study incremental learning for collaboration-aware tokenizers in generative recommendation, and formulate a drift-aware selective adaptation problem that aims to update drifted items while keeping stationary items stable.
    
    \item We propose \textbf{DACT}, a two-stage framework that combines drift identification and selective token adaptation. In particular, we introduce an end-to-end CDIM to estimate item-level drift and enable differentiated updates for drifting and stationary items.
    
    \item We evaluate DACT on three real-world datasets with two representative GRMs, and report improved performance over strong incremental-update baselines under the same setting.
\end{itemize}

\section{Preliminaries}
We follow recent generative recommendation frameworks with a two-stage pipeline~\cite{wang2024learnable,shi2025incremental}: a collaborative tokenizer maps items into discrete tokens, and a generative recommender performs autoregressive prediction over these token sequences. 
\subsection{Collaboration-aware Tokenizer}
\label{sec:pre:catok}
\noindent\textbf{RQ-VAE Tokenizer.}
We implement the tokenizer as an $L$-level residual-quantization VAE (RQ-VAE). For item $i$, we extract static semantic information such as title and description, and obtain a semantic embedding $\mathbf{z}_i \in \mathbb{R}^{d}$ using a
pre-trained content encoder~\cite{rajput2023recommender}. We then feed $\mathbf{z}_i$ into a learnable item encoder $E(\cdot)$, which produces a latent representation
$\mathbf{r}_i=\mathrm{Encoder}(\mathbf{z}_i)$.

RQ-VAE assigns tokens at each level by looking up a codebook $\mathcal{C}^{l}=\{\mathbf{e}^{l}_m\}_{m=1}^{M}$, where  $M$ is the codebook size and $\mathbf{e}^{l}$ is the code embedding at level $l$.
Let $\mathbf{v}_{i,1}=\mathbf{r}_i$ be the initial residual. At level $l$, we define the probability of choosing token $m$ as
\begin{equation}
p(m \mid \mathbf{v}_{i,l})=
\frac{\exp\!\left(-\|\mathbf{v}_{i,l}-\mathbf{e}^{l}_m\|^2/T\right)}
{\sum_{j=1}^{M}\exp\!\left(-\|\mathbf{v}_{i,l}-\mathbf{e}^{l}_j\|^2/T\right)},
\label{prob}
\end{equation}
where $T$ is the temperature. Then select the token index by
\begin{equation}
c_{i,l}=\arg\max_{m} \; p(m \mid \mathbf{v}_{i,l}), \qquad 
\mathbf{v}_{i,l+1}=\mathbf{v}_{i,l}-\mathbf{e}^{l}_{c_{i,l}},
\end{equation}
where $c_{i,l}$ is the $l$-th assigned token for item $i$ and $\mathbf{v}_{i,l+1}$ is the residual vector at the
$l$-th level. Thus each item can be mapped to a hierarchical token sequence $\mathbf{c}_i=[c_{i,1},\ldots,c_{i,L}]$ and the corresponding quantized embedding 
\begin{equation}
\hat{\mathbf{r}}_i=\sum_{l=1}^{L}\mathbf{e}^{l}_{c_{i,l}}.
\label{quant}
\end{equation}

A decoder reconstructs the input representation as $\hat{\mathbf{z}}_i=\mathrm{Decoder}(\hat{\mathbf{r}}_i)$, and we optimize the reconstruction loss
\begin{equation}
\mathcal{L}_{\text{recon}}=\|\mathbf{z}_i-\hat{\mathbf{z}}_i\|^2.
\end{equation}

To update the codebook and stabilize the encoder output, we adopt the standard quantization loss:
\begin{equation}
\mathcal{L}_{\text{rq}} = \sum_{l=1}^{L} \left( \|\mathrm{sg}[\mathbf{v}_{i,l}]-\mathbf{e}^{l}_{c_{i,l}}\|^2 + \mu \|\mathbf{v}_{i,l}-\mathrm{sg}[\mathbf{e}^{l}_{c_{i,l}}]\|^2 \right),
\end{equation}
where $\mathrm{sg}[\cdot]$ denotes the stop-gradient operator and $\mu$ is a hyperparameter.

\noindent\textbf{Collaborative signals injection.}
To incorporate collaborative signals from a well-trained CF recommender model (e.g., SASRec~\cite{kang2018self} and LightGCN~\cite{he2020lightgcn}) into the code sequence, existing collaboration-aware tokenizers~\cite{zheng2025universal,zhou2025onerec,wang2024learnable,xiao2025unger,lin2025cemg,luo2025qarm} commonly introduce an auxiliary \emph{collaborative alignment} objective during tokenizer training, which acts on the tokenizer encoder and encourages the learned latent representation to capture both content and interaction signals. Formally, given an item $i$ and its collaborative embedding $\mathbf{h}_i$, the collaborative alignment loss can be written as
\begin{equation}
\mathcal{L}_{\text{cf}} = f({\mathbf{z}}_i,\mathbf{h}_i),
\end{equation}
where $f(\cdot)$ can be instantiated in various ways (e.g., co-occurring item alignment~\cite{zheng2025universal} or regularization~\cite{wang2024learnable}). Following LETTER~\cite{wang2024learnable}, we obtain CF embeddings for items and align them with the tokenizer's quantized embedding $\hat{\mathbf{r}_i}$ using a collaborative regularization objective:
\begin{equation}
\mathcal{L}_{\text{cf}}=
-\log \frac{\exp(\mathrm{sim}(\hat{\mathbf{r}}_i,\mathbf{h}_i))}
{\sum_{j\in \mathcal{B}}\exp(\mathrm{sim}(\hat{\mathbf{r}}_j,\mathbf{h}_i))},
\end{equation}
where $j$ ranges over items in the mini-batch $\mathcal{B}$ and $\mathrm{sim}(\cdot,\cdot)$ denotes cosine similarity. The overall tokenizer objective is
\begin{equation}
\left\{
\begin{aligned}
\mathcal{L}_{\text{RQ-VAE}} &= \mathcal{L}_{\text{recon}}+\mathcal{L}_{\text{rq}}, \\
\mathcal{L}_{\text{Tokenizer}} &= \mathcal{L}_{\text{RQ-VAE}}+\lambda\,\mathcal{L}_{\text{cf}}
.
\end{aligned}
\right.
\end{equation}
where $\lambda$ controls the strength of collaborative regularization.

\subsection{Generative Recommender Model}
\label{sec:pre:grm}

Given a user $u$ with interaction history $\mathcal{S}_u=[i_1,\ldots,i_T]$, the tokenizer maps each item $i_t$ to a discrete token sequence and we concatenate them as the input token sequence $\mathbf{x}_u=[x_{u,1},\ldots,x_{u,|\mathbf{x}_u|}]$. 
A generative recommender models next-token prediction via autoregressive factorization:
\begin{equation}
p(\mathbf{x}_u)=\prod_{t=1}^{|\mathbf{x}_u|} p(x_{u,t}\mid x_{u,<t}),
\end{equation}
and the next token is predicted by
\begin{equation}
\hat{x}_{u,t}=\arg\max_{v\in\mathcal{V}} \; p(v \mid x_{u,<t}),
\end{equation}
where $\mathcal{V}$ is the token vocabulary. During training, we minimize the negative log-likelihood of the ground-truth next tokens:
\begin{equation}
\mathcal{L}_{\text{NLL}}
= - \sum_{u\in\mathcal{U}} \sum_{t=1}^{|\mathbf{x}_u|}
\log p(x_{u,t}\mid x_{u,<t}).
\label{nextloss}
\end{equation}

\subsection{Continuous Learning for GRM}
With new items and new interactions arriving over time , continuous learning is essential for GRM. We denote the data stream as ${\{\mathcal{D}_0,\mathcal{D}_1,\dots ,\mathcal{D}_p,\dots\}}$, where $p$ denotes the $p$-th time period. We denote the model parameters at period $p$ as $M_p$, where $M_0$ represents the parameters of the model pre-trained on $\mathcal{D}_0$. 
When performing incremental training at time period $p$, we initialize the model parameters with $M_{p-1}$ and update them using the data from $\mathcal{D}_p$.

\section{Methods}
\begin{figure*}[t]
    \centering
    \includegraphics[width=\textwidth]{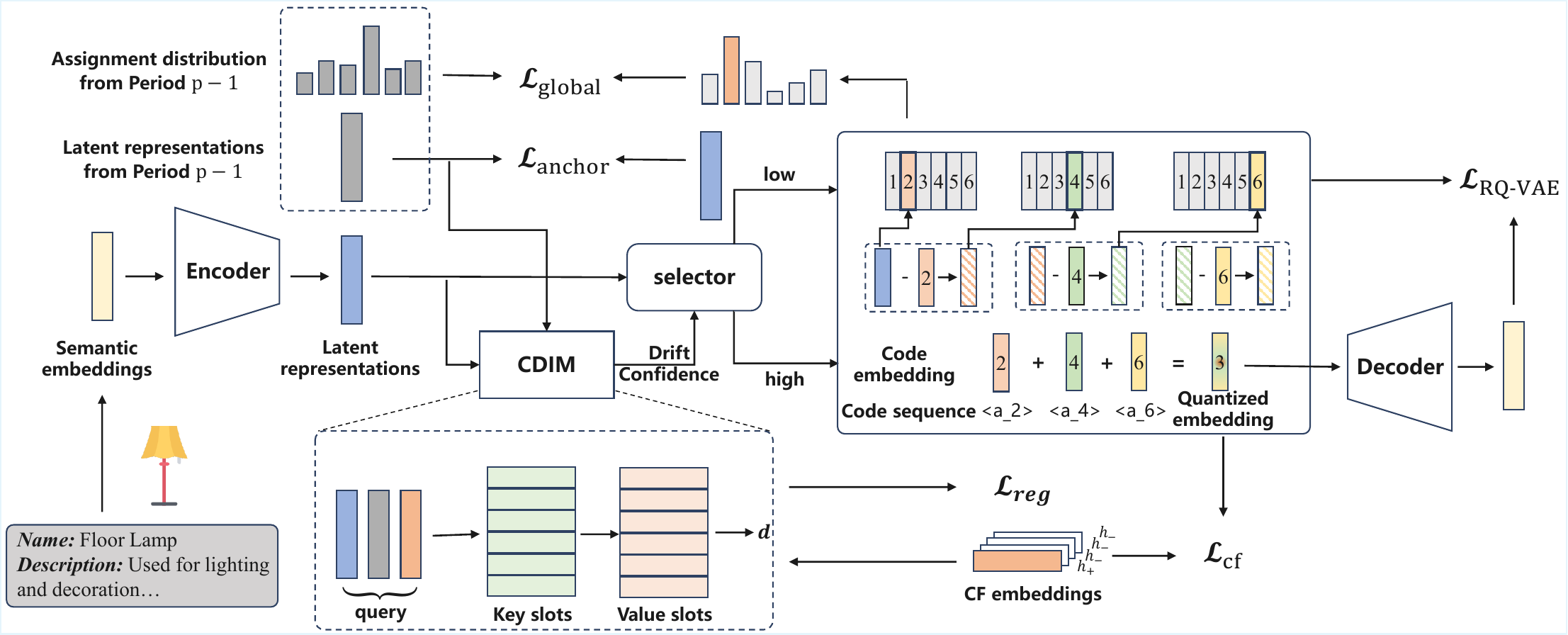}
    \caption{The framework of drift-aware tokenizer adaptation in DACT. DACT introduces CDIM to learn drift patterns and update-policy prototypes, which predicts a drift confidence score to guide the differentiated training strategy. Additionally, a global code-assignment stability constraint is applied to all items.}
    \label{fig:framework}
\end{figure*}

In our framework, we decompose the item token update procedure into two stages. In the first stage, to obtain latent representations aligned with evolving collaborative signals, we introduce a CDIM trained end-to-end with the tokenizer, as illustrated in Fig. ~\ref{fig:framework}. In the second stage, we perform hierarchical code reassignment based on the fine-tuned latent representations from Stage 1.

\subsection{Drift-Aware Tokenizer Adaptation }
\label{sec:method:inc}

\subsubsection{Collaborative Drift Identification Module}
\label{sec:method:cdim}

To capture item-level collaborative drift patterns, we introduce the CDIM, which is jointly trained with the tokenizer during fine-tuning. CDIM learns a reusable set of drift patterns and update-policy prototypes. For each item, CDIM predicts a drift confidence score $d_i$. It takes three inputs: (i) the latent representation of item $i$ from the previous period $p\!-\!1$, denoted as $\mathbf{r}^{p-1}_i$; (ii) its current latent representation during training , denoted as $\tilde{\mathbf{r}}^{p}_i$; and (iii) collaborative embedding $\mathbf{h}^{p}_i$ from fine-tuned CF recommender models in period $p$.
 For newly arrived items, we compute $\mathbf{r}^{p-1}$ by feeding their semantic embeddings into the previous-stage tokenizer encoder at period $p-1$. It compares how $\mathbf{r}^{p-1}_i$ and $\tilde{\mathbf{r}}^{p}_i$ match the new collaborative signal, together with the latent representation change, and then queries the most relevant drift pattern to make an update decision. During training, CDIM is only used to guide the \emph{differentiated} update strategy and thus is \emph{not} needed at inference time.

\noindent\textbf{Pattern memory.}
We maintain two learnable slots: key slots $\mathbf{K}\in\mathbb{R}^{S\times D}$ representing drift patterns, and value slots $\mathbf{V}\in\mathbb{R}^{S\times D}$ representing the corresponding update policies, where $S$ is the number of slots and $D$ is the hidden dimension. At the first continual update period ($p=1$), we randomly initialize the pattern memory $\{\mathbf{K},\mathbf{V}\}$.  For subsequent periods ($p>1$), we warm-start CDIM by loading $\{\mathbf{K}^{p-1},\mathbf{V}^{p-1}\}$ from the previous period $p-1$, and continue optimizing them end-to-end together with the tokenizer.

\noindent\textbf{Query construction.}
We measure the alignment between the previous-period and current-period latent representations and the latest collaborative embedding, and use their difference as change information.
\begin{equation}
\mathbf{s}^{p-1}_i=\mathbf{r}^{p-1}_i\odot \mathbf{h}^{p}_i,\quad
\tilde{\mathbf{s}}^{p}_i=\tilde{\mathbf{r}}^{p}_i\odot \mathbf{h}^{p}_i,\quad
\Delta\tilde{\mathbf{r}}^{p}_i=\tilde{\mathbf{r}}^{p}_i-\mathbf{r}^{p-1}_i,
\end{equation}
where $\odot$ denotes element-wise product. We then concatenate them to form the query vector:
\begin{equation}
\mathbf{q}_i=\mathrm{concat}\!\left[\mathbf{s}^{p-1}_i,\tilde{\mathbf{s}}^{p}_i,\Delta\tilde{\mathbf{r}}^{p}_i\right].
\end{equation}
To stabilize training, we stop gradients through $\tilde{\mathbf{r}}^{p}_i$ (i.e., detach it from the computation graph).

\noindent\textbf{Pattern matching and confidence prediction.}
We compute attention weights over key slots:
\begin{equation}
\boldsymbol{\alpha}_i=\mathrm{Softmax}\!\left(\frac{\mathbf{q}_i\mathbf{K}^\top}{\tau}\right),
\label{cdimscore}
\end{equation}
where $\tau$ is a temperature. We aggregate value slots and map the result to a drift confidence:
\begin{equation}
d_i=\sigma(MLP(\boldsymbol{\alpha}_i\mathbf{V})),
\end{equation}
where $\sigma(\cdot)$ is the sigmoid and $d_i\in(0,1)$ indicates the confidence of triggering a code update for item $i$ (higher means more likely drifting, lower means more likely stationary). To prevent a trivial shortcut solution with uninformative confidences, we add a lightweight regularizer on the confidence:
\begin{equation}
\mathcal{L}_{\text{reg}}=\frac{1}{|\mathcal{I}|}\sum_{i\in\mathcal{I}} d_i^2,
\end{equation}
where $\mathcal{I}$ denotes the set of items and $|\mathcal{I}|$ is the number of items.

\subsubsection{Differentiated Update Strategy}
\label{sec:method:diff}

Based on the drift confidence $d_i$ predicted by CDIM, we apply differentiated update strategy. Alternative gating schemes such as per-item thresholding can be prone to collapsing toward a single side (e.g., selecting almost all or almost no items), especially under indirect supervision. Therefore, We select the top-$K$ items with the highest confidence as the drifting set $\mathcal{I}_d$, and treat the remaining items as the stationary set $\mathcal{I}_s$:
\begin{equation}
\mathcal{I}_d = \operatorname*{arg\,topK}_{i\in\mathcal{I}} \; d_i,
\qquad
\mathcal{I}_s = \mathcal{I}\setminus \mathcal{I}_d,
\end{equation}
where $K\in(0,1)$ is a ratio hyperparameter that specifies the fraction of items selected as drifting. To enable end-to-end training, we use a straight-through estimator~\cite{bengio2013estimating} for the non-differentiable indexing operation, so that gradients from the downstream losses $\mathcal{L}_{\text{drift}}$ and $\mathcal{L}_{\text{stable}}$ (Eqs.~\ref{newloss} and~\ref{oldloss}) can flow back to CDIM. We then apply differentiated objectives to encourage plasticity for drifting items and stability for stationary items.

\noindent\textbf{Drifting items ($\mathcal{I}_d$).}
For items in $\mathcal{I}_d$, we encourage their latent representations to adapt freely to the latest collaborative signals. Accordingly, we minimize the following objective:
\begin{equation}
\mathcal{L}_{\text{drift}}
=\frac{1}{|\mathcal{I}_d|}\sum_{i\in\mathcal{I}_d}\Big(\mathcal{L}_{\text{RQ-VAE}}(i)+\lambda\,\mathcal{L}_{\text{cf}}(i)\Big).
\label{newloss}
\end{equation}

\noindent\textbf{Stationary items ($\mathcal{I}_s$).}
For items in $\mathcal{I}_s$, we constrain their latent representations to preserve identifier stability and, in turn, prevent disrupting the GRM's learned token-embedding alignment. We introduce an anchoring regularizer:
\begin{equation}
\mathcal{L}_{\text{anchor}}(i)=\left\|\tilde{\mathbf{r}}^{p}_i-\mathbf{r}^{p-1}_i\right\|^2,\qquad i\in\mathcal{I}_s,
\end{equation}
and define the stationary item objective as
\begin{equation}
\mathcal{L}_{\text{stable}}
=\frac{1}{|\mathcal{I}_s|}\sum_{i\in\mathcal{I}_s}\Big(\mathcal{L}_{\text{RQ-VAE}}(i)+\alpha\,\mathcal{L}_{\text{anchor}}(i)+\lambda \mathcal{L}_{\text{cf}}(i)\Big),
\label{oldloss}
\end{equation}
where $\alpha$ controls the strength of anchoring. We retain the CF term for stationary items to keep their representations weakly aligned with the current collaborative space, preventing their latent representations from becoming obsolete.

\subsubsection{Global Code-Assignment Stability Constraint}
\label{sec:method:kl}

To preserve identifier stability, we regularize the token-assignment distribution across all items. Specifically, for each item $i$ and level $l$, we denote the assignment distribution as the $M$-dimensional probability vector
\begin{equation}
\mathbf{p}_{i,l}=\big[p(1\mid \mathbf{v}_{i,l}),\,\ldots,\,p(M\mid \mathbf{v}_{i,l})\big]\in\mathbb{R}^{M},
\end{equation}
where $p(m\mid \mathbf{v}_{i,l})$ is defined in Eq.~\ref{prob}, with temperature $T$ to control the sharpness of the assignment distribution.

We focus on the first codebook layer ($l=1$), and compute the corresponding distributions at periods $p\!-\!1$ and $p$, denoted as $\mathbf{p}^{p-1}_{i,1}$ and $\mathbf{p}^{p}_{i,1}$, respectively. 
We penalize their shift by averaging the KL divergence over all items:
\begin{equation}
\mathcal{L}_{\text{global}}
=\frac{1}{|\mathcal{I}|}\sum_{i\in\mathcal{I}}
D_{\mathrm{KL}}\!\left(\mathbf{p}^{p-1}_{i,1} \,\|\, \mathbf{p}^{p}_{i,1}\right).
\end{equation}

We apply $\mathcal{L}_{\text{global}}$ only to the first layer, since deeper residual layers are more sensitive to quantization noise; moreover, their reassignment is explicitly controlled by our hierarchical strategy in Sec.~\ref{sec:method:assign}.

\subsubsection{Overall Objective}
\label{sec:method:obj}

The overall drift-aware tokenizer adaptation objective is
\begin{equation}
\mathcal{L}_{\text{Tokenizer}}
=\mathcal{L}_{\text{drift}}+\theta\,\mathcal{L}_{\text{stable}}+\beta \,\mathcal{L}_{\text{global}}+\zeta \mathcal{L}_{reg},
\end{equation}
where $\theta$ controls the relative update strength between drifting and stationary items, and $\beta$ controls the strength of the global code-assignment stability constraint. 

For CDIM, under the straight-through estimator, gradients from $\mathcal{L}_{\text{drift}}$, $\mathcal{L}_{\text{stable}}$, and $\mathcal{L}_{\text{reg}}$ jointly shape the confidence scores. Intuitively, assigning a truly drifting item to the stationary set incurs a stronger optimization tension between the anchoring term and alignment to the latest collaborative signals, which is typically harder to minimize; in contrast, genuinely stationary items tend to exhibit lower collaborative mismatch and can better balance the anchoring constraint. As a result, CDIM is encouraged to assign higher drift confidence to drifting items and lower confidence to stationary ones, leading to a better overall objective.

\subsection{Code Reassignment Strategy}
\label{sec:method:assign}

Due to the residual-quantization nature of RQ-VAE, later codebook layers quantize progressively smaller residuals and therefore tend to encode finer-grained information. Meanwhile, to maximally preserve the token–embedding alignment learned by the GRM at stage $p!-!1$, we aim to minimize unnecessary token changes. Accordingly, we adopt a relaxed-to-strict strategy:

\noindent\textbf{Layer-1: Active Adaptation.}
For the first layer, we always reassign codes by running inference with the fine-tuned tokenizer at stage $p$:
\begin{equation}
c^{p}_{i,1} \leftarrow \arg\max_{k}\; p\!\left(k \mid \mathbf{v}^{p}_{i,1}\right),
\end{equation}
where $\mathbf{v}^{p}_{i,1}=\mathbf{r}^{p}_i$. This is because the first-layer token captures the dominant semantic cluster. Any change here indicates a significant collaborative drift that must be reflected immediately.

\noindent\textbf{Deeper-layers: Conditional Stability.}
For deeper layers ($l\ge 2$), we apply a conditional update rule. We trigger reassignment \textit{if and only if} the first-layer code changes ($c^{p}_{i,1} \neq c^{p-1}_{i,1}$). Otherwise, we force the subsequent tokens to remain unchanged: $c^{p}_{i,l} \leftarrow c^{p-1}_{i,l}$.
This design filters out minor fluctuations in the latent space, thereby stabilizing the majority of identifiers.

After reassignment, we obtain the updated token sequences for each item in period $p$. We then fine-tune the GRM for period $p$ on the evolving interaction data using these updated tokens as item identifiers, by minimizing the next-token prediction loss in Eq.~\ref{nextloss}, which balances stability and plasticity.


\section{Experiment}
In this section, we conduct extensive experiments on three real-world datasets with two representative GRMs to answer the following research questions:
 \textbf{RQ1:} How does DACT perform compared with different baselines?
 \textbf{RQ2:} How do the key components of DACT affect performance?
 \textbf{RQ3:} How sensitive is DACT to different hyperparameter settings?
 \textbf{RQ4:} Compared with standard continuous learning methods, how efficient is DACT in terms of training speed?

\subsection{Experimental Settings}
\begin{table}[t]
\centering
\small
\setlength{\tabcolsep}{6pt}
\caption{Statistics of the datasets.}
\label{tab:stat}
\begin{tabular}{lcccc}
\toprule
Dataset & \#Users & \#Items & \#Interactions & Sparsity \\
\midrule
Beauty & 22,363 & 12,101 & 198,502 & 99.927\% \\
Tools  & 16,638 & 10,217 & 134,476 & 99.921\% \\
Toys   & 19,412 & 11,924 & 167,597 & 99.928\% \\
\bottomrule
\end{tabular}
\end{table}
\subsubsection{Datasets}
We use three real-world datasets spanning different domains: \textbf{Beauty}, \textbf{Tools}, and \textbf{Toys} from Amazon.
\begin{table*}[t]
\centering
\caption{Overall performance of baselines and DACT on TIGER. The best results are highlighted in \textbf{bold}, and the second-best results are \underline{underlined}. "Tok." is the abbreviation for the tokenizer, while "$\times$", "FT", "RT" refer to "Frozen", "Fine-tuning" and "Retraining" respectively.}
\label{tab:tiger}
\resizebox{\textwidth}{!}{
\begin{tabular}{l|cc|cccc|cccc|cccc|cccc}
\hline 
\multirow{2}{*}{\textbf{Dataset}} & \multicolumn{2}{c|}{\textbf{Method}} & \multicolumn{4}{c|}{\textbf{Period 1}} & \multicolumn{4}{c|}{\textbf{Period 2}} & \multicolumn{4}{c|}{\textbf{Period 3}} & \multicolumn{4}{c}{\textbf{Period 4}} \\
\cline{2-19} 
& Tok. & GRM & H@5 & H@10 & N@5 & N@10 & H@5 & H@10 & N@5 & N@10 & H@5 & H@10 & N@5 & N@10 & H@5 & H@10 & N@5 & N@10 \\
\hline 
\multirow{6}{*}{Beauty} 
 & $\times$ & $\times$ & 0.0279 & 0.0443 & 0.0151 & 0.0192 & 0.0207 & 0.0341 & 0.0105 & 0.0139 & 0.0172 & 0.0268 & 0.0086 & 0.0111 & 0.0147 & 0.0223 & 0.0070 & 0.0089 \\
 & FT & $\times$ & 0.0065 & 0.0104 & 0.0035 & 0.0045 & 0.0049 & 0.0106 & 0.0020 & 0.0034 & 0.0026 & 0.0044 & 0.0011 & 0.0016 & 0.0041 & 0.0092 & 0.0019 & 0.0031 \\
& $\times$ & FT & 0.0274 & 0.0469 & 0.0162 & 0.0226 & 0.0211 & 0.0407 & 0.0128 & 0.0191 & 0.0293 & 0.0524 & 0.0185 & 0.0259 & \underline{0.0336} & 0.0514 & \underline{0.0207} & 0.0265 \\
& FT & FT & 0.0270 & \underline{0.0483} & 0.0169 & \underline{0.0237} & 0.0218 & 0.0422 & 0.0129 & 0.0194 & {0.0328} & {0.0547} & 0.0200 & 0.0270 & 0.0316 & \underline{0.0575} & 0.0205 & \underline{0.0288} \\
& FT & RT & 0.0253 & 0.0401 & 0.0156 & 0.0203 & 0.0208 & 0.0378 & 0.0125 & 0.0180 & \underline{0.0392} & \textbf{0.0628} & \underline{0.0245} & \textbf{0.0321} & 0.0321 & 0.0534 & 0.0198 & 0.0267 \\
& \multicolumn{2}{c|}{Reformer} & \underline{0.0285} & 0.0483 & \underline{0.0171} & 0.0235 & \underline{0.0246} & \underline{0.0438} & \underline{0.0153} & \underline{0.0214} & 0.0328 & 0.0544 & 0.0212 & 0.0281 & 0.0285 & 0.0515 & 0.0189 & 0.0263 \\
\cline{2-19}
& \multicolumn{2}{c|}{DACT} & \textbf{0.0299} & \textbf{0.0531} & \textbf{0.0191} & \textbf{0.0266} & \textbf{0.0267} & \textbf{0.0465} & \textbf{0.0167} & \textbf{0.0230} & \textbf{0.0398} & \underline{0.0605} & \textbf{0.0250} & \underline{0.0316} & \textbf{0.0368} & \textbf{0.0609} & \textbf{0.0228} & \textbf{0.0305} \\
\hline 

\multirow{6}{*}{Tools} 
 & $\times$ & $\times$ & 0.0201 & 0.0283 & 0.0103 & 0.0123 & 0.0188 & 0.0285 & 0.0102 & 0.0126 & 0.0146 & 0.0227 & 0.0080 & 0.0101 & 0.0128 & 0.0191 & 0.0069 & 0.0084 \\
 & FT & $\times$ & 0.0045 & 0.0067 & 0.0019 & 0.0025 & 0.0017 & 0.0054 & 0.0007 & 0.0017 & 0.0025 & 0.0050 & 0.0013 & 0.0019 & 0.0021 & 0.0034 & 0.0011 & 0.0014 \\
& $\times$ & FT & 0.0175 & 0.0330 & 0.0113 & 0.0163 & \underline{0.0224} & 0.0312 & 0.0132 & 0.0160 & \underline{0.0246} & 0.0322 & \underline{0.0162} & 0.0187 & 0.0178 & 0.0310 & 0.0110 & 0.0152 \\
& FT & FT & 0.0192 & 0.0347 & 0.0114 & 0.0164 & 0.0196 & 0.0315 & 0.0123 & 0.0162 & 0.0214 & \underline{0.0372} & 0.0138 & \underline{0.0188} & {0.0205} & \underline{0.0317} & \underline{0.0130} & \underline{0.0166} \\
& FT & RT & 0.0134 & 0.0231 & 0.0082 & 0.0113 & 0.0174 & 0.0291 & 0.0117 & 0.0155 & 0.0158 & 0.0230 & 0.0101 & 0.0125 & \underline{0.0209} & 0.0312 & 0.0128 & 0.0161 \\
& \multicolumn{2}{c|}{Reformer} & \underline{0.0231} & \underline{0.0360} & \underline{0.0143} & \underline{0.0184} & {0.0213} & \underline{0.0317} & \underline{0.0141} & \underline{0.0175} & {0.0234} & 0.0354 & {0.0146} & 0.0186 & {0.0205} & 0.0314 & 0.0124 & 0.0158 \\
\cline{2-19}
& \multicolumn{2}{c|}{DACT} & \textbf{0.0239} & \textbf{0.0414} & \textbf{0.01453} & \textbf{0.0201} & \textbf{0.0233} & \textbf{0.0355} & \textbf{0.0149} & \textbf{0.0187} & \textbf{0.0255} & \textbf{0.0380} & \textbf{0.0164} & \textbf{0.0204} & \textbf{0.0246} & \textbf{0.0374} & \textbf{0.0151} & \textbf{0.0192} \\
\hline 

\multirow{6}{*}{Toys} 
 & $\times$ & $\times$ & 0.0184 & 0.0303 & 0.0101 & 0.0132 & 0.0154 & 0.0229 & 0.0083 & 0.0102 & 0.0185 & 0.0278 & 0.0096 & 0.0119 & 0.0132 & 0.0250 & 0.0070 & 0.0010 \\
 & FT & $\times$ & 0.0059 & 0.0135 & 0.0025 & 0.0044 & 0.0048 & 0.0089 & 0.0025 & 0.0035 & 0.0055 & 0.0116 & 0.0024 & 0.0040 & 0.0075 & 0.0135 & 0.0029 & 0.0045 \\
& $\times$ & FT & 0.0159 & 0.0260 & 0.0103 & 0.0136 & 0.0152 & 0.0272 & 0.0092 & 0.0130 & 0.0282 & 0.0398 & 0.0176 & 0.0214 & 0.0235 & 0.0398 & 0.0154 & 0.0206 \\
& FT & FT & 0.0176 & 0.0305 & 0.0117 & 0.0159 & \underline{0.0192} & \textbf{0.0326} & \underline{0.0116} & \textbf{0.0159} & \underline{0.0284} & \underline{0.0456} & \underline{0.0185} & \underline{0.0240} & \underline{0.0262} & \underline{0.0430} & \underline{0.0169} & \underline{0.0223} \\
& FT & RT & 0.0176 & \underline{0.0329} & 0.0120 & \underline{0.0169} & 0.0159 & 0.0252 & 0.0100 & 0.0130 & 0.0196 & 0.0311 & 0.0116 & 0.0153 & 0.0248 & 0.0389 & 0.0160 & 0.0205 \\
& \multicolumn{2}{c|}{Reformer} & \underline{0.0188} & 0.0301 & \underline{0.0127} & 0.0163 & 0.0176 & 0.0266 & {0.0110} & 0.0139 & 0.0240 & 0.0382 & 0.0154 & 0.0200 & 0.0232 & 0.0373 & 0.0156 & 0.0201 \\
\cline{2-19}
& \multicolumn{2}{c|}{DACT} & \textbf{0.0208} & \textbf{0.0353} & \textbf{0.0140} & \textbf{0.0186} & \textbf{0.0200} & \underline{0.0318} & \textbf{0.0117} & \underline{0.0155} & \textbf{0.0309} & \textbf{0.0523} & \textbf{0.0197} & \textbf{0.0267} & \textbf{0.0269} & \textbf{0.0455} & \textbf{0.0179} & \textbf{0.0239} \\
\hline 
\end{tabular}%
}
\end{table*}
Beauty contains user interactions with a large number of beauty products; Toys covers diverse toys and game items; and Tools records user behaviors on hardware tools and home improvement equipment.
Following prior work~\cite{liu2025improving, shi2025incremental}, we preprocess the data by filtering out users and items with fewer than 5 interactions. We sort interactions chronologically and obtain period-wise datasets $\{\mathcal{D}_0,\ldots,\mathcal{D}_4\}$. We refer to these periods as $P0,\ldots,P4$, where $\mathcal{D}_0$ (period $P0$) contains the first 60\% interactions for pre-training and $\mathcal{D}_1,\ldots,\mathcal{D}_4$ (periods $P1$--$P4$) equally partition the remaining 40\% interactions for continual updates~\cite{shi2024preliminary}.
We construct user sequences for training and evaluation using a sliding-window strategy~\cite{yoo2025continual}, and split the data into training/validation/test sets using the leave-one-out protocol.
The statistics of users, items, and sparsity are reported in Table~\ref{tab:stat}.

\subsubsection{Backbones and Baseline}

We deploy DACT on two representative generative recommendation backbones. For \textbf{TIGER~\cite{rajput2023recommender}}, following prior work, we adopt a T5-based encoder--decoder architecture as the GRM and fine-tune it in a standard way. For \textbf{LCRec~\cite{zheng2024adapting}}, we fine-tune a decoder-only LLM initialized from Qwen2.5-1.5B-Instruct following ~\cite{shi2025incremental}, using LoRA for parameter-efficient training; in each period, we load the LoRA weights from the previous period for continual updates.

During incremental learning, the tokenizer and the GRM can be either continuously fine-tuned or kept frozen, leading to following baselines: 
(1) \textbf{Both frozen}, where the tokenizer and the GRM are fixed; 
(2) \textbf{Fine-tuning tokenizer} while GRM is frozen; 
(3) \textbf{Fine-tuning GRM} while GRM is incrementally updated; 
(4) \textbf{Both fine-tuning};
(5) \textbf{Reformer~\cite{shi2025incremental}}, which applies the tokenizer-side incremental strategy to allocate identifiers for newly arrived items. 
In addition, for \textbf{TIGER}, we further compare: 
(6) \textbf{Fine-tuning tokenizer and retraining GRM};
For \textbf{LCRec}, since it is trained with LoRA in a continual manner, full retraining is prohibitively expensive; instead, we compare two LoRA-based continual recommendation baselines: 
(7) \textbf{LSAT}, which retrains a LoRA module to capture short-term preferences while cumulatively updating another LoRA module for long-term preferences; and 
(8) \textbf{PESO}, which adds a proximal regularizer on LoRA updates to balance stability and plasticity.

For fine-tuning, at period $p$, each model is initialized from the checkpoint trained on $\mathcal{D}_{p-1}$ and is trained only on data from $\mathcal{D}_{p}$; for retraining, the model is initialized from scratch and trained on all data accumulated up to period $p$, i.e., $\bigcup_{t=0}^{p}\mathcal{D}_{t}$. Similarly, the latest collaborative signals are obtained from a CF model fine-tuned on $\mathcal{D}_{p}$.

\subsubsection{Implementation Details}

For the tokenizer in the initial stage, we use an RQ-VAE with a 3-layer codebook with 256 code embeddings of 32 dimensions in each layer, and train it with AdamW for 20K steps. Following prior works~\cite{rajput2023recommender, wang2024learnable}, we set learning rate as 1e-3, $\mu$ as 0.25 and $\lambda$ as 0.02. For the continuous stage of DACT, we train for 5,000 steps with a learning rate of 1e-4 and $\zeta$ as 0.001, $S$ as 32, $D$ as 64, and search $\alpha$ in a range of \{0.5, 1.0, 2.0, 5.0\}, $\theta$ in a range of \{0.001, 0.1, 0.5, 1.0\}, $\beta$ in a range of \{0.5, 1.0, 5, 10\}, $K$ in a range of\{0.1, 0.3, 0.7, 0.9\}.

\subsubsection{Evaluation Metrics}
We evaluate recommendation performance using two widely adopted HR@k (H@k) and NDCG@k (N@k)~\cite{liu2025improving}. Following prior work~\cite{wang2024learnable,rajput2023recommender,shi2025incremental}, we set $k=5$ and 10.

\begin{table*}[t]
\centering
\caption{Overall performance of baselines and DACT on LC-Rec.}
\label{tab:lcrec}
\resizebox{\textwidth}{!}{%
\begin{tabular}{l|cc|cccc|cccc|cccc|cccc}
\hline
\multirow{2}{*}{\textbf{Dataset}} & \multicolumn{2}{c|}{\textbf{Method}} & \multicolumn{4}{c|}{\textbf{Period 1}} & \multicolumn{4}{c|}{\textbf{Period 2}} & \multicolumn{4}{c|}{\textbf{Period 3}} & \multicolumn{4}{c}{\textbf{Period 4}} \\
\cline{2-19}
 & Tok. & GRM & H@5 & H@10 & N@5 & N@10 & H@5 & H@10 & N@5 & N@10 & H@5 & H@10 & N@5 & N@10 & H@5 & H@10 & N@5 & N@10 \\
\hline
\multirow{8}{*}{Beauty}
 & $\times$  & $\times$  & 0.0136 & 0.0237 & 0.0080 & 0.0112 & 0.0096 & 0.0186 & 0.0055 & 0.0084 & 0.0049 & 0.0129 & 0.0027 & 0.0053 & 0.0079 & 0.0141 & 0.0050 & 0.0069 \\
 & FT & $\times$  & 0.0042 & 0.0064 & 0.0022 & 0.0028 & 0.0038 & 0.0069 & 0.0022 & 0.0032 & 0.0011 & 0.0023 & 0.0006 & 0.0010 & 0.0021 & 0.0036 & 0.0011 & 0.0016 \\
 & $\times$  & FT & 0.0228 & 0.0394 & 0.0134 & 0.0188 & 0.0100 & 0.0198 & 0.0058 & 0.0089 & 0.0136 & 0.0257 & 0.0073 & 0.0111 & 0.0183 & 0.0334 & 0.0109 & 0.0157 \\
 & FT & FT & \underline{0.0229} & 0.0420 & \underline{0.0139} & \underline{0.0200} & \underline{0.0215} & \underline{0.0431} & \underline{0.0125} & \underline{0.0194} & \underline{0.0245} & \underline{0.0486} & \underline{0.0148} & \underline{0.0225} & \textbf{0.0319} & \underline{0.0545} & \textbf{0.0194} & \underline{0.0267} \\
 
 & \multicolumn{2}{c|}{LSAT} & 0.0177 & 0.0294 & 0.0109 & 0.0147 & 0.0106 & 0.0198 & 0.0065 & 0.0102 & 0.0106 & 0.0219 & 0.0065 & 0.00102 & 0.0172 & 0.0314 & 0.0110 & 0.0155 \\
 & \multicolumn{2}{c|}{PESO} & 0.0211 & 0.0347 & 0.0128 & 0.0171 & 0.0183 & 0.0351 & 0.0110 & 0.0163 & 0.0109 & 0.0256 & 0.0063 & 0.0110 & 0.0214 & 0.0362 & 0.0125 & 0.0172 \\
 & \multicolumn{2}{c|}{Reformer} & 0.0215 & \underline{0.0428} & 0.0119 & 0.0187 & 0.0100 & 0.0257 & 0.0047 & 0.0096 & 0.0202 & 0.0389 & 0.0111 & 0.0170 & 0.0272 & 0.0455 & 0.0166 & 0.0224 \\
\cline{2-19}
 & \multicolumn{2}{c|}{DACT} & \textbf{0.0262} & \textbf{0.0477} & \textbf{0.0157} & \textbf{0.0225} & \textbf{0.0239} & \textbf{0.0440} & \textbf{0.0144} & \textbf{0.0209} & \textbf{0.0270} & \textbf{0.0546} & \textbf{0.0165} & \textbf{0.0253} & \underline{0.0300} & \textbf{0.0576} & \underline{0.0189} & \textbf{0.0279} \\
\hline
\multirow{8}{*}{Tools}
 & $\times$  & $\times$  & 0.0125 & 0.0207 & 0.0082 & 0.0108 & 0.0112 & 0.0203 & 0.0074 & 0.0103 & 0.0095 & 0.0169 & 0.0059 & 0.0083 & 0.0109 & 0.0173 & 0.0068 & 0.0089 \\
 & FT & $\times$  & 0.0038 & 0.0062 & 0.0024 & 0.0031 & 0.0033 & 0.0058 & 0.0021 & 0.0029 & 0.0024 & 0.0042 & 0.0015 & 0.0020 & 0.0027 & 0.0046 & 0.0017 & 0.0023 \\
 & $\times$  & FT & 0.0157 & 0.0269 & 0.0102 & 0.0138 & 0.0147 & 0.0246 & 0.0093 & 0.0124 & 0.0110 & 0.0194 & 0.0061 & 0.0088 & 0.0050 & 0.0123 & 0.0028 & 0.0052 \\
 & FT & FT & \underline{0.0214} & \underline{0.0310} & \underline{0.0134} & \underline{0.0165} & {0.0162} & \underline{0.0276} & \underline{0.0100} & \underline{0.0136} & \textbf{0.0218} & \underline{0.0333} & \textbf{0.0146} & \underline{0.0183} & \underline{0.0214} & \textbf{0.0360} & \underline{0.0134} & \underline{0.0181} \\
 & \multicolumn{2}{c|}{LSAT} & 0.0153 & 0.0259 & 0.0088 & 0.0123 & 0.0146 & 0.0237 & 0.0083 & 0.0113 & 0.0121 & 0.0189 & 0.0098 & 0.0125 & 0.0157 & 0.0264 & 0.0101 & 0.0135 \\
 & \multicolumn{2}{c|}{PESO} & 0.0183 & 0.0285 & 0.0116 & 0.0149 & \underline{0.0162} & 0.0244 & 0.0103 & 0.0130 & 0.0097 & 0.0193 & 0.0059 & 0.0089 & 0.0160 & 0.0287 & 0.0100 & 0.0140 \\
 & \multicolumn{2}{c|}{Reformer} & 0.0164 & 0.0300 & 0.0102 & 0.0146 & 0.0114 & 0.0205 & 0.0072 & 0.0101 & 0.0121 & 0.0223 & 0.0067 & 0.0100 & 0.0130 & 0.0275 & 0.0070 & 0.0117 \\
\cline{2-19}
 & \multicolumn{2}{c|}{DACT} & \textbf{0.0226} & \textbf{0.0364} & \textbf{0.0152} & \textbf{0.0196} & \textbf{0.0181} & \textbf{0.0325} & \textbf{0.0109} & \textbf{0.0155} & \underline{0.0209} & \textbf{0.0372} & \underline{0.0136} & \textbf{0.0188} & \textbf{0.0246} & \underline{0.0355} & \textbf{0.0147} & \textbf{0.0181} \\
\hline
\end{tabular}%
}
\end{table*}

\subsection{Overall Performance(RQ1)}
\subsubsection{Performance on TIGER}
\label{sec:exp:tiger:r1}
The comparison between our method and the baselines on the TIGER backbone is reported in Table~\ref{tab:tiger}, from which we observe that:
\begin{itemize}[leftmargin=*]

\item Freezing both the tokenizer and GRM leads to consistent degradation from Period~1 to Period~4, confirming continual preference evolution. Fine-tuning only the tokenizer can even cause sharp drops, as naive updates may trigger widespread token reassignment and break identifier consistency, undermining the GRM's learned token--embedding alignment.

\item Incrementally updating both the tokenizer and GRM generally outperforms updating only the GRM, since the tokenizer must reflect newly introduced collaborative signals (e.g., evolving popularity and co-occurrence). With a fixed tokenizer, the GRM is trained on stale token sequences that no longer match current item semantics.

\item Reformer improves over a frozen tokenizer by adapting codes for newly arrived items. It can be competitive in early periods by keeping old-item codes fixed and preserving token--embedding alignment, but falls behind over time as fixed identifiers fail to track collaborative drift of existing items.

\item DACT consistently performs best by jointly handling (i) code assignment for new items and (ii) selective code updates for drifting items, while constraining stationary identifiers to preserve token--embedding alignment and avoid disruptive token shifts. In some periods, fine-tuning tokenizer and retraining GRM can be stronger, likely because that period requires longer-term preference consolidation that benefits from retraining the GRM on accumulated data.

\end{itemize}

\subsubsection{Performance on LC-Rec}

The overall performance comparison on the LCRec backbone is reported in Table~\ref{tab:lcrec}. Compared with the TIGER results, we have the following observations:
\begin{itemize}[leftmargin=*]
    \item Consistent with TIGER, on the Qwen2.5-1.5B-Instruct-based LC-Rec backbone, freezing the GRM causes a steady performance drop from Period~1 to Period~4 on both datasets.

\item Reformer degrades faster on LC-Rec as periods progress. A plausible reason is that LLM-based GRMs can overfit to the collaborative semantics encoded in token sequences; when signals drift, outdated tokens become misleading priors, leading to larger performance drops and underscoring the need to update existing identifiers.

\item LSAT and PESO often outperform fine-tuning the GRM alone, likely due to their emphasis on long/short-term preferences and stability, respectively. However, with a frozen tokenizer (similar to Reformer), identifiers cannot reflect evolving collaborative signals, so their gains diminish as drift accumulates.

\item DACT consistently matches or outperforms both fine-tuning baselines by balancing plasticity and stability: it updates tokens for drifting items to expose current collaborative signals, enabling more effective learning by the LLM-based GRM.

\end{itemize}

\subsection{Ablation Study(RQ2)}
To thoroughly examine the role of each component in DACT, we conduct an ablation study on the \textbf{Tools} dataset with the \textbf{TIGER} backbone. Specifically, we take the variant that fine-tunes both the tokenizer and the GRM as our reference setting, and then progressively remove key designs in DACT: (1) removing CDIM and randomly selecting items for differentiated update strategy (w/o CDIM); (2) removing the differentiated update strategy (w/o Diff); (3) removing global code-assignment stability constraint (w/o Global); and (4) removing the code re-assignment strategy (w/o Re-assign). 

\begin{figure}[t]
    \centering
    \begin{subfigure}[b]{0.48\columnwidth}
        \centering
        \includegraphics[width=\linewidth]{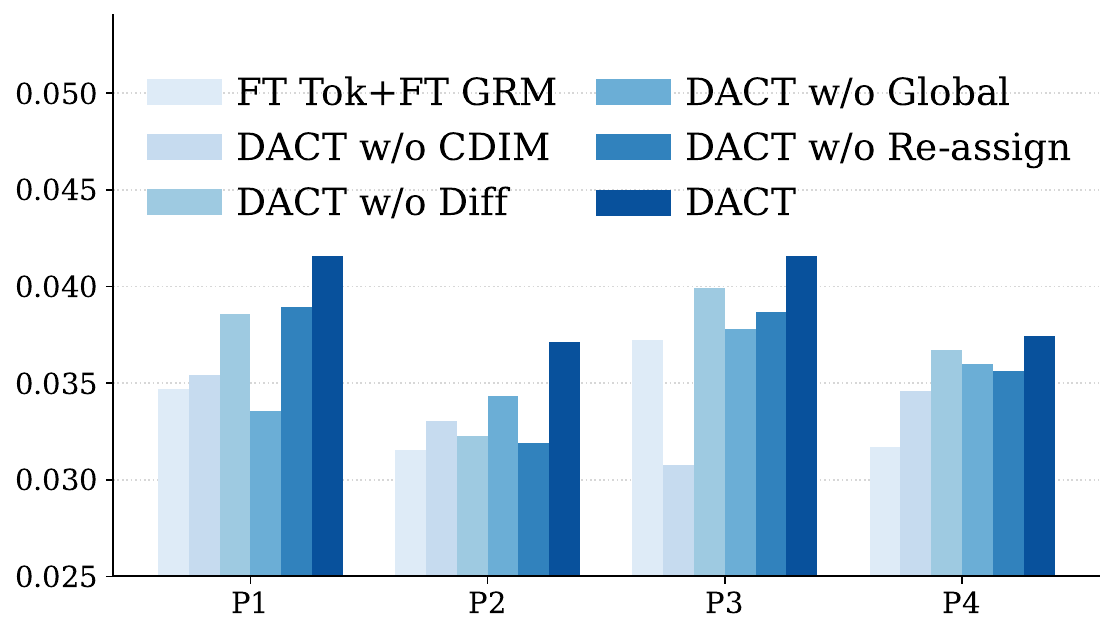}
        \caption{H@10}
        \label{fig:ablation_hr}
    \end{subfigure}
    \hfill 
    \begin{subfigure}[b]{0.48\columnwidth}
        \centering
        \includegraphics[width=\linewidth]{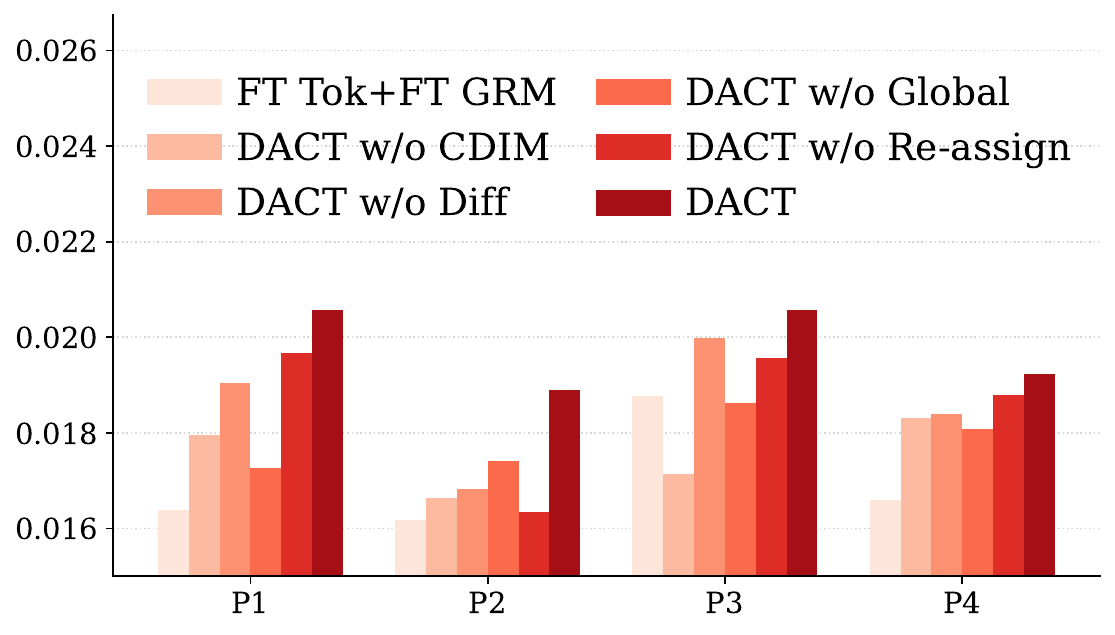}
        \caption{N@10}
        \label{fig:ablation_ndcg}
    \end{subfigure}
    
    \caption{Ablation study on the Tools dataset (TIGER). }
    \label{fig:ablation_main}
\end{figure}

Figure~\ref{fig:ablation_main} shows that DACT achieves the best performance, validating the necessity of each component. Removing CDIM causes a large drop, indicating that effective differentiated updates require accurate drift identification. Both the differentiated update strategy and the global stability constraint improve over the reference, suggesting they are complementary in balancing plasticity and stability. Finally, removing the reassignment strategy degrades performance, confirming the benefit of layer-wise differentiated token reassignment.

\subsection{Further Analysis}
\subsubsection{plasticity: Adaptation to Drift(RQ2)}

\begin{figure}[t]
    \centering
    \includegraphics[width=\columnwidth]{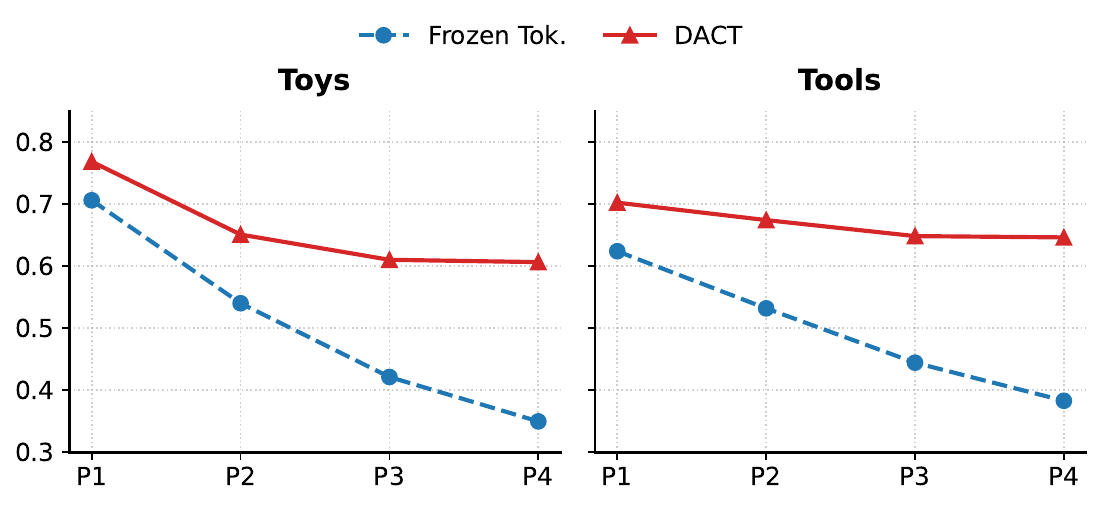}
    \caption{Average cosine similarity between quantized embeddings and CF embeddings on Toys and Tools.  }
    \label{fig:c2}
\end{figure}

We further investigate whether DACT can perceive the evolution of the collaborative landscape and accordingly fine-tune the tokenizer to adapt to collaborative signal drift over time. To quantify the capability of the tokenizer in capturing evolving collaborative signals, we measure the alignment between the \emph{quantized embeddings} from Eq. ~\ref{quant} and the \emph{CF embeddings} of the current period. Specifically, for each period, we compute the cosine similarity between the two representations for every item, and report the average over all items. We consider two tokenizer settings: a frozen tokenizer and our DACT-updated tokenizer. The results are reported in Figure~\ref{fig:c2}.

For the frozen tokenizer (Frozen Tok.), similarity steadily declines over time, reflecting \emph{collaborative drift} as the initial semantic space becomes increasingly misaligned with evolving collaborative signals. In contrast, DACT maintains relatively stable similarity across periods, indicating better adaptation to the latest collaborative patterns.

\begin{figure}[t]
    \centering
    \begin{subfigure}[b]{0.48\columnwidth}
        \centering
        \includegraphics[width=\linewidth]{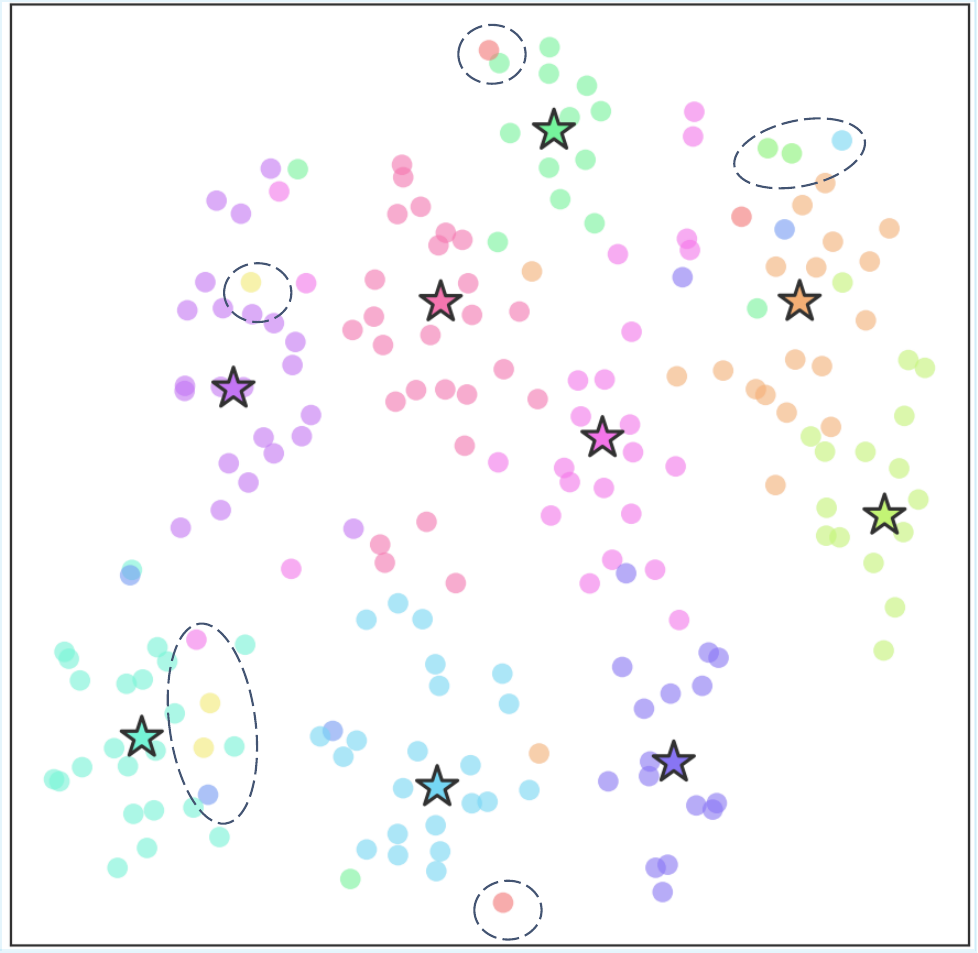}
        \caption{Frozen Tok.}
    \end{subfigure}
    \hfill 
    \begin{subfigure}[b]{0.48\columnwidth}
        \centering
        \includegraphics[width=\linewidth]{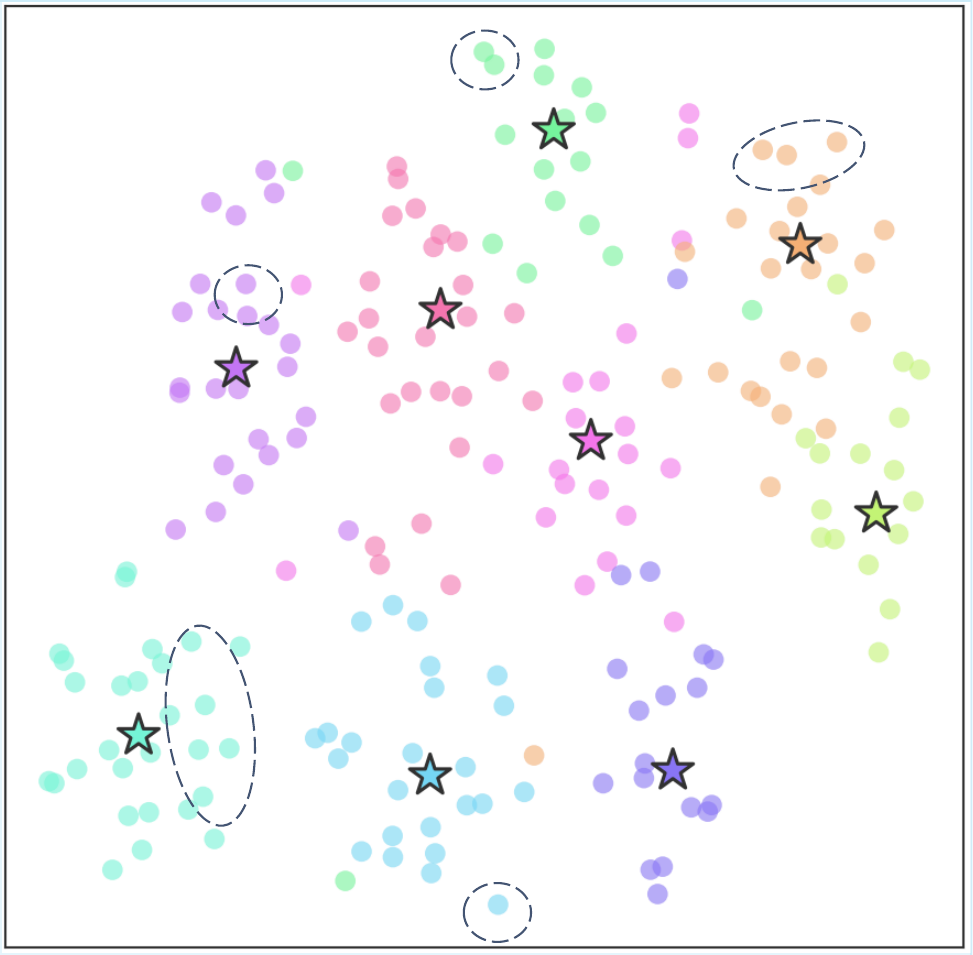}
        \caption{DACT}
    \end{subfigure}
    
    \caption{Visualization of Period 1 CF embeddings from Tools, where colors denote first-layer token assignments and stars denote code embeddings.}
    \label{fig:assign}
\end{figure}

To intuitively understand how DACT handles specific instances of drift, we selected 200 items from Tools and visualized their CF embeddings in Period 1 using t-SNE~\cite{tsne}. The items are colored according to their assigned tokens in the first layer of the codebook. As shown in Figure ~\ref{fig:assign}, we visualize the token assignments in Period 1 using (a) a frozen tokenizer inherited from Period 0 and (b) the tokenizer updated by DACT, and the corresponding code embeddings are marked by stars. In both subfigures, most items lie close to the code embedding of their assigned token, suggesting that the first-layer tokens generally capture the dominant collaborative signals. However, in Fig.~\ref{fig:assign}(a), the items highlighted by circles deviate from their original code embedding and become closer to other code embeddings, indicating that their collaborative semantics have drifted in Period 1. In contrast, Fig.~\ref{fig:assign}(b) shows that DACT reassigns these drifted items to tokens whose code embeddings better match their current CF embeddings, thereby adapting the identifiers to the latest collaborative signals.

\subsubsection{Stability: Changes of Code Assignment(RQ2)}
To examine the stability of our method, we measure the average code-assignment change rate of each layer across periods and compare it with naively fine-tuning the tokenizer. Here, Overall indicates whether an item changes its code in at least one layer. The results are reported in Table~\ref{tab:ablation_change_rate}.

\begin{table}[h]
    \centering
    \caption{Average code change rates across layers for different methods on the Tools dataset.}
    \label{tab:ablation_change_rate}
    \begin{tabular}{lcccc}
        \toprule
        \textbf{Method} & \textbf{Layer 1} & \textbf{Layer 2} & \textbf{Layer 3}  & \textbf{Overall} \\
        \midrule
        Fine-tuning         & 0.7052 & 0.9696 & 0.9892 & 0.9999\\
        \midrule
        DACT ($K=0.1$)      & 0.1331 & 0.1237 & 0.1300 & 0.1331\\
        DACT ($K=0.3$)      & 0.2975 & 0.2812 & 0.2918 & 0.2975\\
        DACT ($K=0.7$)      & 0.4592 & 0.4435 & 0.4524 & 0.4592\\
        \midrule
        DACT ($\beta=10$)   & 0.2200 & 0.2065 & 0.2157 & 0.2200\\
        DACT ($\beta=5$)    & 0.2975 & 0.2812 & 0.2918 & 0.2975\\
        DACT ($\beta=0.5$)  & 0.4091 & 0.3902 & 0.4012 & 0.4091\\
        \bottomrule
    \end{tabular}
\end{table}

From Table~\ref{tab:ablation_change_rate}, DACT yields substantially lower code change rates than full fine-tuning across all layers, whereas fine-tuning triggers large-scale reassignment and near-complete changes in deeper layers (Layer 2/3). Besides, we find that smaller $K$ or larger $\beta$ further reduces change rates, leading to more stable code assignments.

\subsubsection{Investigation on Warm/Cold items(RQ2)}
\begin{figure}[t]
    \centering
    \includegraphics[width=\columnwidth]{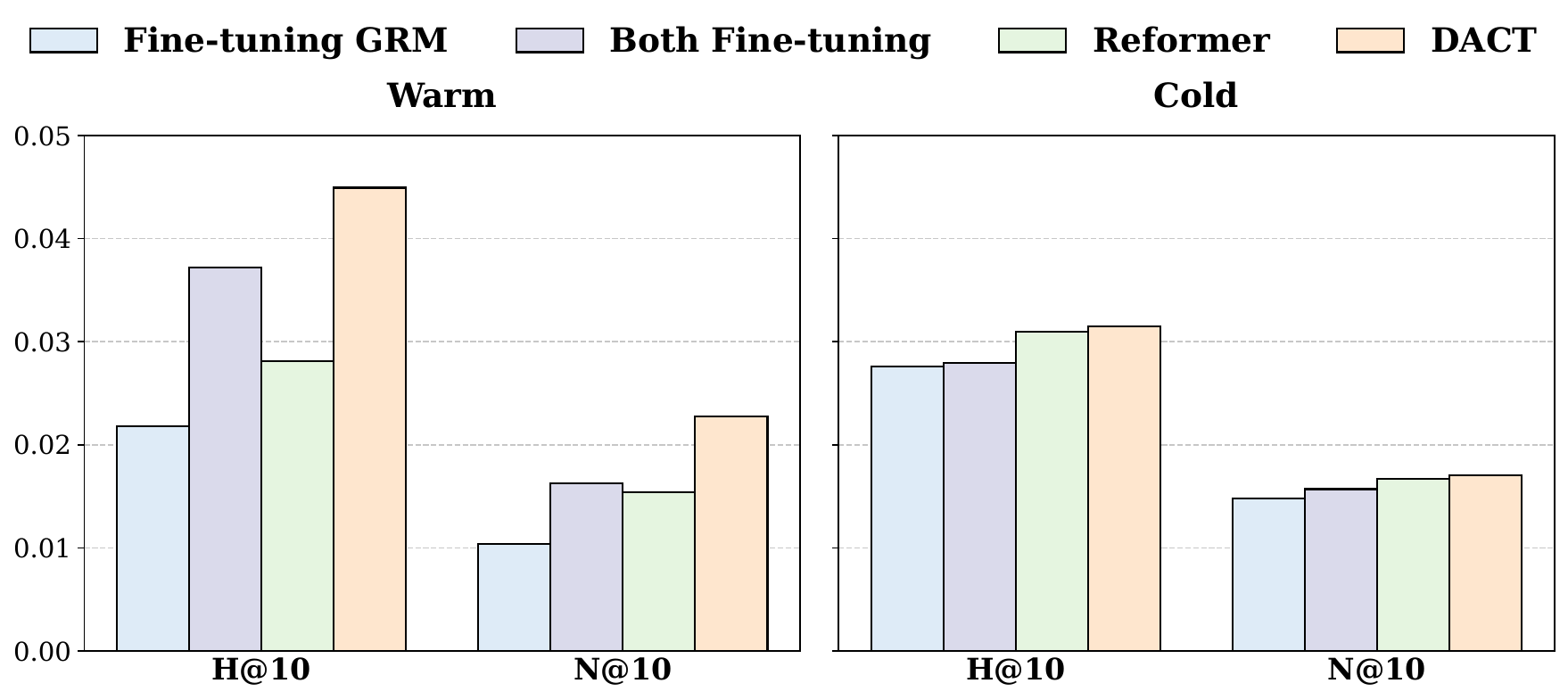}
    \caption{Warm/Cold item performance of TIGER on Toys across different methods.}
    \label{fig:cold}
\end{figure}

To examine DACT's capability of adapting to newly emerged items, we partition items into \emph{warm} and \emph{cold} sets following~\cite{shi2025incremental}. 
Specifically, items that appear in the training data of the tokenizer at period $P0$ are regarded as warm items, while items that first appear after $P0$ are treated as cold items. We evaluate TIGER on the Toys dataset and report results on period $P1$. Figure~\ref{fig:cold} compares DACT with three competitive baselines on warm and cold items, respectively. We can find Reformer performs better than both fine-tuning and only fine-tuning tokenizer on cold items, indicating its effectiveness in assigning identifiers to newly emerged items. However, Reformer underperforms joint tokenizer fine-tuning on warm items, highlighting the necessity of adapting to newly arriving collaborative signals. By considering both newly emerged items and new interactions, our method achieves the best overall performance.

\subsubsection{Hyper-Parameter Analysis(RQ3)}
\label{sec:exp:hparam}
\begin{figure*}[t]
    \centering
    \includegraphics[width=\textwidth]{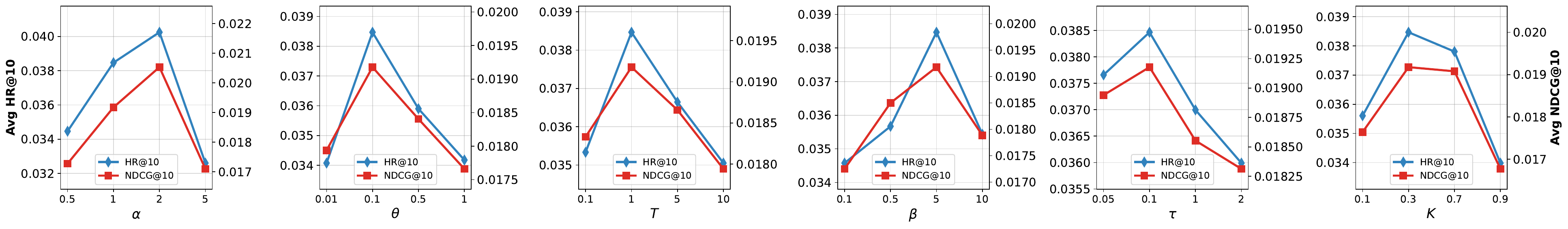}
    \caption{Performance of DACT based on TIGER over different hyper-parameters on Tools}
    \label{fig:ab}
\end{figure*}

We analyze key hyper-parameters of DACT on the Tools dataset with the TIGER backbone. Figure~\ref{fig:ab} reports the average H@10 and N@10 over four periods by varying one parameter at a time.

\begin{itemize}[leftmargin=*, labelwidth=0pt, labelsep=0.6em, align=left]

\item \textbf{Stable loss weight $\theta$.} 
Sweeping $\theta\in[0.01,1]$, the best performance is achieved at $\theta{=}0.1$. Too small $\theta$ weakens stability constraints, while too large $\theta$ overemphasizes stability and suppresses necessary adaptation.

\item  \textbf{Temperature $T$ in global loss.}
The global loss regularizes the shift of the assignment \emph{distribution} across periods, which enforces not only consistent code selection but also consistent relative distances to all code embeddings. We vary $T$ from 0.1 to 10, and a moderate value $T=1$ performs best.

\item  \textbf{Global loss weight $\beta$.}
$\beta$ controls how strongly we constrain the first-layer assignment distribution. Larger $\beta$ yields fewer code changes but weaker adaptation; smaller $\beta$ allows more changes but less stability. Varying $\beta$ from 0.1 to 10, we find $\beta=5$ works best.


\item \textbf{Temperature $\tau$ in CDIM.}
By varying $\tau$ from 0.05 to 2, we observe that an overly large $\tau$ makes the attention scores in Eq.~\ref{cdimscore} overly smooth, causing different drift patterns to receive similar weights and thus weakening CDIM’s ability to distinguish drifting items.

\item \textbf{Top selection $K$ in CDIM.}
$K$ controls the split between drifting and stationary items, and thus directly affects the relative contributions of $\mathcal{L}{\text{drift}}$ and $\mathcal{L}{\text{stable}}$ during training. Sweeping $K$ from 0.1 to 0.9, we observe a rise-then-fall trend: extreme values (e.g., $K{=}0.1$ or $K{=}0.9$) make almost all items dominated by one group, which largely nullifies the differentiated update strategy and leads to noticeable degradation. In contrast, a moderate ratio $K{=}0.3$ achieves the best performance.
\end{itemize}

\subsubsection{Time Efficiency Analysis(RQ4)}
\begin{figure}[t]
    \centering
    \begin{subfigure}[b]{0.45\columnwidth}
        \centering
        \includegraphics[width=\linewidth]{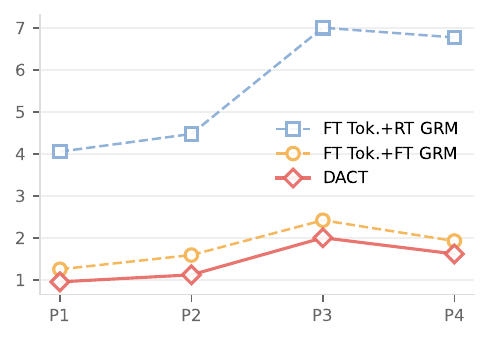}
        \caption{Toys}
    \end{subfigure}
    \hfill 
    \begin{subfigure}[b]{0.45\columnwidth}
        \centering
        \includegraphics[width=\linewidth]{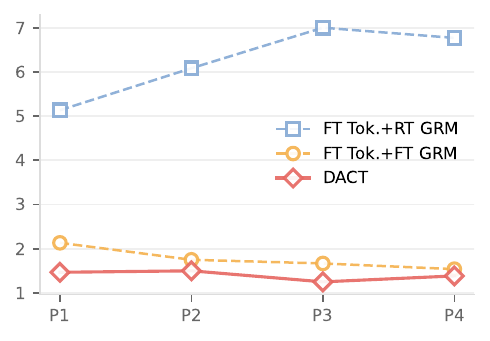}
        \caption{Tools}
    \end{subfigure}
    
    \caption{Time consumption ratio of different continual learning methods on TIGER, normalized by the time of only fine-tuning GRM, across two datasets.}
    \label{exp:time}
\end{figure}

We evaluate the training time consumption of the GRM under different updating strategies. Taking the most efficient baseline—Fine-tune GRM without Tokenizer updates—as the unit of measurement ($1.0$), we calculate the relative training time ratios for other methods across different periods, as illustrated in Figure  \ref{exp:time}. The results demonstrate that our proposed DACT consistently achieves superior efficiency compared to strategies that involve joint Fine-tuning of both the Tokenizer and the GRM. Meanwhile, retraining the GRM incurs substantially higher time cost than fine-tuning, and is therefore typically impractical in real-world settings.

\section{Related Work}
\subsection{Generative Recommendation}


Generative recommendation formulates recommendation as sequence generation~\cite{li2024survey,wang2023generative,hou2025generating,vuong2025mechanisms}. Early work such as P5 leverages LLMs to unify multiple tasks~\cite{geng2022recommendation}, while TIGER~\cite{rajput2023recommender} and LETTER~\cite{wang2024learnable} introduce learnable codebooks (e.g., RQ-VAE) to map items into short token sequences, alleviating long textual identifiers and large ID vocabularies~\cite{hua2023index}. Beyond compression, recent methods~\cite{xiao2025unger,wang2024content,wang2024learnable,wang2024eager,zhou2025onerec,Wang2025} incorporate collaborative signals via co-occurrence, regularization, or contrastive objectives, yielding substantial gains.

In real systems, however, interactions evolve continuously, calling for incremental learning~\cite{yongjiwu2021rethinking,caroprese2025modelling}. Reformer~\cite{shi2025incremental} assigns codes to newly arrived items and PESO~\cite{yoo2025continual} explores parameter-efficient updates, but they typically assume the collaborative semantics of previously tokenized items are static, overlooking drift from popularity shifts, seasonality, and other temporal effects. When an old item’s interaction pattern changes, keeping its code fixed can create semantic conflicts in the token space. We explicitly study this issue and propose a drift-aware incremental codebook adaptation mechanism.

\subsection{Continuous Learning}
In practice, interactions arrive continuously and new items emerge, making incremental training indispensable under limited computation and latency budgets~\cite{huang2025training,liu2023recommendation,chen2023continual,liu2023autoseqrec}. A core challenge is the stability--plasticity trade-off~\cite{wang2024comprehensive,liu2025agentcf++}: retaining prior knowledge while adapting to new data. Although incremental learning has been widely studied in recommender systems, generative recommendation introduces additional difficulty. Many GRMs adopt a decoupled tokenizer--recommender architecture, yet couple them through learned codes: stability requires existing token embeddings to remain valid, while plasticity demands timely adaptation to drift from evolving interactions.

Recent methods explore efficient updates, e.g., LSAT (dual LoRA for long/short-term interests)~\cite{shi2024preliminary}, PESO (constraint-based retention)~\cite{yoo2025continual}, and Reformer (incremental coding for new items)~\cite{shi2025incremental}. However, they largely overlook code updates for \emph{existing} items under collaborative drift. In contrast, we selectively update codes for drifting items while constraining stationary ones, improving both effectiveness and efficiency.

\section{Conclusion}
In this work, we address collaborative drift in generative recommendation with DACT, a framework that continually updates tokenizers while preserving previously learned token–embedding alignments. The key idea is to separate “what should change” from “what should stay stable” when item representations evolve over time. To this end, DACT combines (i) a Collaborative Drift Identification Module (CDIM) that detects drifting versus stationary items and applies differentiated fine-tuning to avoid over-updating stable regions, and (ii) a hierarchical code reassignment strategy that constrains remapping to only the necessary subset of tokens, thereby reducing disruptive code switches and maintaining semantic consistency across stages. As a result, DACT improves both adaptation to new interactions and retention of prior knowledge, offering a practical continual tokenization solution for generative recommendation. Extensive experiments on three datasets and multiple evaluation settings demonstrate that DACT consistently outperforms strong baselines, validating its effectiveness under collaborative drift.

For future work, promising directions include the following. First, building on CF-only tokenization schemes~\cite{lin2025order}, it is interesting to explore alternative continual learning strategies under decoupled collaborative signals. Second, for large-scale item corpora, it remains to be studied whether hierarchical reassignment rules should vary across codebook layers and scales.

\bibliography{ref.bib}
\end{document}